# Redefining the Attraction Measure, Scaling Exponent and Impedance Function of the Gravity Model

Yanguang Chen[1], Feng Xu[2]

(1. Department of Geography, Peking University, Beijing 100871, China; 2. China Center for Town Reform and Development, Beijing 100045, China)

**Abstract:** The attraction measure, scaling exponent, and impedance function of the gravity model are redefined using the concepts from fractals and spatial complexity. Firstly, the attraction measure of spatial interaction in human systems is defined by the product of traffic inflow and outflow. Based on the new definition, the gravity model originating from the Newtonian analogy is differentiated from Wilson's spatial interaction model deriving from entropy-maximizing principle. Secondly, the scaling exponent of the gravity model based on the inverse distance relationship is revealed to be the ratio of the fractal dimension of networks of cities to that of hierarchies of cities. The value of the scaling exponent is then shown to approach to 2, which corresponds to the value of the power exponent of the law of gravity in classical physics. Thirdly, the inverse power function is demonstrated to be more acceptable than the negative exponential function as an impedance function of the gravity model. The limits of application of different gravity models based on different impedance functions are brought to light. Applying the expanded theory and models to China's interregional flows yields satisfying results, which in turn lend further support to the expanded theory and models.

**Key words:** spatial interaction, the gravity model, impedance function, scaling exponent, fractal dimension

# 1 Introduction

The ideas from system theory play an important role in geographical research. Cities and regions are both complex social spatial systems (Allen, 1997; Wilson, 2000). In order to effectively research into social and economic development in a region, we should have the systematical concepts. Spatial interaction is the precondition of the existence of urban and regional systems. Without interaction there would be no urban and regional systems. One of the acceptable models for quantificationally describing spatial interaction is the gravity model. There are lots of theoretical and empirical explorations of spatial interaction associated with the gravity models (Erlander, 1980; Fotheringham and O'Kelly, 1989; Haynes and Fotheringham, 1984; Lierop, 1986; Roy, 2004; Sen and Smith, 1995). Nevertheless, a mass of researches do not imply a broad space for developing the spatial interaction theory of economic- and social-geographical systems. Recently, the conventional gravity model of interaction between cities is improved by introducing a time variable and a lag parameter to the attraction measures and using Fourier analysis and cross-correlation function (Chen and Wang, 2008). The expanded model can be applied to analyzing spatial-temporal process of aggregate human behaviors related to spatial interaction.

However, some basic questions of the gravity model such as the attraction measures, the model parameters, and the impedance functions are still pending for further discussion. How to measure the attractive force between any two places? What is the mathematical essence of the distance friction coefficient? Which type of function, the inverse power function or the negative exponential function, is more reasonable for characterizing the distance impedance? All these problems remain to be solved. This paper is just devoted to redefining the attractive measure, distance exponent and impedance function of the gravity model.

The remaining part of this paper is structured as following. Section 2 introduces the general gravity model and discusses the selections of different models. The attraction measure of spatial interaction is redefined and the scaling exponent indicative of distance friction is shown to be a ratio of fractal dimensions. Section 3 shows how to apply the expanded theory and models to China's interregional flows by railway, and explains the modeling process including data sources, technical route, and detailed calculation results. Section 4 discusses the related basic theoretical issues, which give some new lights on the selection and application of different gravity models in human geography. Finally, this paper is concluded with a brief summary.

## 2 Model selections

### 2.1 General form of the gravity model

The gravity model is often expressed in different forms for different purposes in different branches of human geography. The general gravity model of spatial interaction between places $i$ and $j$ can be written as

$$F_{ij} = G P_i^\alpha P_j^\beta f(r_{ij}), \tag{1}$$

where the interaction volume between two places $F_{ij}$ is positively related to their sizes $P_i$ and $P_j$ and inversely related to the distance $r_{ij}$ between them. $G$, $\alpha$, and $\beta$ are parameters. $G$ is the gravity coefficient. In particular, $f(\cdot)$ is the impedance function, which is always measured in two forms: one is an inverse power function, and the other is a negative exponential function. If we choose the inverse distance relationship as a distance impedance, then equation (1) can be rewritten as

$$F_{ij} = G P_i^\alpha P_j^\beta r_{ij}^{-\gamma}. \tag{2}$$

where $\gamma$ is the distance friction coefficient. According to equation (2), spatial interaction has the effect of action as a distance, namely interaction will not totally attenuate as distance increases. If we choose the negative exponential function for impedance, then equation (1) can be rewritten as

$$F_{ij} = G P_i^\alpha P_j^\beta e^{-\lambda r_{ij}}. \tag{3}$$

where $\lambda$ is the distance-decay coefficient. According to equation (3), spatial interaction has the effect of locality (Chen, 2008), namely interaction will be close to 0 when distance increases to a certain value ($r_{ij}>1/\lambda$).

Each of the two models shown above has its advantages and disadvantages. The gravity model based on the inverse power function, namely equation (2), is derived from classical physics by analogy with the Newton's law of universal gravitation. Although it can well reflect the effect of action at a distance, it has two weaknesses: (1) its underlying rationale seems to be uncertain; (2) there are some problems in its dimensional explanation according to the traditional concepts of geometry (Haynes, 1975). By contrast, the gravity model based on the negative exponential function, namely equation (3), has a clear theoretical base and no dimensional dilemma. The exponential impedance function can be deduced from entropy maximization principle—all the negative exponential models can be derived by using entropy-maximizing methods in theory (Chen, 2008). However, equation (3) is weak in reflecting the effect of action at a distance. Owing to the development of fractal theory and its application to human geography (Batty and Longley, 1994; Frankhauser, 1994; Mandelbrot, 1983), the dimensional problem of equation (2) can be resolved today. Further researches indicate that there is no strict confliction between equations (2) and (3) in essence. As a matter of fact, each model has its own sphere of application.

## 2.2 Redefinition of attraction measure

A specific difficult to apply the gravity models to spatial interaction is the definition of the attraction measurement. What variable should be employed to measure gravity? For a long time, most geographers adopt interurban or interregional traffic flow as the attraction measurement, which is quite acceptable. However, in both theory and practice, it is not the most proper way to use traffic flow as attraction force directly. One possible hypothesis is that attraction measure is a function of traffic flow rather than traffic flow itself. For simplicity, we might as well assume that the parameters $\alpha$, $\beta$ in equation (1) are equal to each other under certain condition. Then we have

$$I_{ij} = F_{ij}^{1/\alpha} = G^{1/\alpha} P_i P_j f(r_{ij})^{1/\alpha} = K P_i P_j g(r_{ij}), \qquad (4)$$

where

$$I_{ij} = F_{ij}^{1/\alpha}, K = G^{1/\alpha}, g(r_{ij}) = f(r_{ij})^{1/\alpha}.$$

As we know, traffic flows includes inflows associated with origins ($O_{ij}$) and outflow with destinations ($D_{ij}$). Generally speaking, inflow is not equal to outflow ($O_{ij} \neq D_i$), i.e., they have an asymmetric relation. In this case, $\alpha$ does not equal $\beta$. The solution to this problem is to take the sum or product of inflow and outflow to replace a single flow. We know that both addition and multiplication satisfy the commutative properties. Consequently, both of them can enable the redefined flow measures to satisfy symmetrical requirement. And yet, the traffic flow measure based on addition has different dimensions compared with the product of size ($P_i P_j$), while the flow measure based on multiplication is of dimensional consistency with the size product. So, we define the flow measure in the form

$$F_{ij} = O_{ij} * D_{ji}, \qquad (5)$$

The attraction measure could be further redefined as

$$I_{ij} = (O_{ij}D_{ji})^{1/\alpha}. \tag{6}$$

By doing so, the parameter $\alpha$ becomes equal to another parameter $\beta$. Thus equation (2) can be reexpressed in the form

$$I_{ij} = KP_iP_jr_{ij}^{-b}, \tag{7}$$

in which the distance friction parameter $b=\gamma/\alpha$ is in fact a distance scaling exponent. Correspondingly, equation (3) can be reexpressed as

$$I_{ij} = KP_iP_je^{-br_{ij}}, \tag{8}$$

where the distance-decay parameter $b=\gamma/\alpha$ is still a distance-decay coefficient.

The negative exponential function based gravity model, equation (3), is identical in form to the Wilson's spatial interaction model derived from the principle of entropy-maximization (Wilson, 1970; Wilson, 2000). Just because of this, equation (3) is always confused with the Wilson's model. However, the gravity model and the Wilson's model reflect different but relational spatial dynamics of human systems. Wilson's model is defined by $O_iD_jf(r_{ij})$, where $O_i = \sum_j O_{ij}$, $D_j = \sum_i D_{ji}$, and the function is used to predict the flow $O_{ij}$ or $D_{ji}$. The gravity model is defined by $P_iP_jf(r_{ij})$, as shown by equation (8), and the formula is employed to evaluate the attraction volume $I_{ij}=f(O_i, D_j)$. Suppose that there are allometric relationships between the size measures ($P_i$, $P_j$) and flow measures ($O_i$, $D_j$). The total flows can be defined as $O_i \propto P_i^{\alpha}$, $D_i \propto P_j^{\beta}$, where $\alpha$ and $\beta$ denote allometric scaling exponents asscociated with fractal dimension. Only in this instance, the Wilson's spatial interaction model can be reconstrtucted and transformed into the exponential-based gravity model, equation (3).

## 2.3 Derivation of scaling exponent

As a gravity model, equation (7) seems to be more reasonable and thus acceptable than equation (8). First, equation (7) bears an analogy with the Newton's law of gravitation. Second, equation (7) can reflect action at a distance of spatial dynamics. Third, equation (7) can be used to explain the spatial complexity of human geographical systems. Therefore, it is necessary to research into equation (7), especially, to bring to light the nature of the scaling exponent $b$.

The key of studying geographical laws is to make use of hierarchical structure. There is a mathematical relationship between a hierarchy and a network (Batty and Longley, 1994; Chen and Zhou, 2006). We can explore the dynamical mechanism of a network through a corresponding hierarchy and vice versa. Human geographical systems such as network of cities always take on hierarchical structure. For a hierarchy of cities with cascade structure, we have the following scaling relation (Chen and Zhou, 2008)

$$f_m = \mu P_m^{-D}, \qquad (9)$$

where $f_m$ is the city number of order $m$, $P_m$ is the corresponding average population size of the same order, the constant $\mu = f_1 P_1^D$ is the proportionality coefficient, and the exponent $D = 1/q$ is the fractal dimension of the urban hierarchy. Here $q$ is the power of Zipf's law, and it can be called "Zipf's dimension" (Chen and Zhou, 2003).

Suppose there exists a balance point $k$ between place $i$ and place $j$. According to equation (7), we get

$$I_{ik} = K P_i P_k r_{ik}^{-b} = K P_j P_k r_{jk}^{-b} = I_{jk}, \qquad (10)$$

Then a complex scaling relation can be derived from equation (10) as follows

$$\frac{P_i}{P_j} = (\frac{r_{ik}}{r_{jk}})^b, \qquad (11)$$

which relates to Reilly's law of retail gravitation (Reilly, 1931). Applying equation (11) to hierarchy of cities yields

$$\frac{P_m}{P_{m+1}} = (\frac{L_m}{L_{m+1}})^b, \qquad (12)$$

in which $L=2r$ is the distance between two central places of order $m$. Thus we have

$$P_m = \eta L_m^b, \qquad (13)$$

where $\eta = P_1 L_1^{-b}$. Substituting equation (13) into equation (9) yields

$$f_m = \mu \eta^{-D} L_m^{-Db} = \varphi L_m^{-D_f}, \qquad (14)$$

in which $\varphi = \mu \eta^{-1/q} = f_1 L_1^{D_f}$, $D_f = Db$, and $D_f$ is the fractal dimension of network of cities (Chen and Zhou, 2006). For the standard rank-size distribution of cities, we have $D = 1/q = 1$ (Chen and Zhou, 2003). On the other hand, the fractal dimension $D_f = Db$ approach to 2 (i.e. $D_f \to 2$) in theory if the geographical space of the ideal earth's surface is fully filled up by cities (Chen and Zhou, 2006).

So the scaling exponent of the gravity model based on the inverse distance relationship is

$$b = \frac{D_f}{D} = qD_f \to 2, \quad (15)$$

where the notation "→" denotes "approach to". This suggests that the scaling exponent of the gravity model can be determined by the ratio of the fractal dimension of network of cities to that of hierarchies of cities. The fractal dimension of hierarchies of cities is actually equal to that of city-size distribution (Chen and Zhou, 2003; Chen and Zhou, 2008). In practice, the value of $b$ is always around 2, and if the spatial sample is large enough, the $b$ value will be very close to 2. The following part will demonstrate the fact that the gravity analysis based on China's interregional flows verifies the above definition of attraction measure and the value of the scaling exponent.

## 3 Empirical evidence—Modelling China's interregional flows

### 3.1 Data sources and methods

The precondition of selecting the right geographical gravity model and estimating its parameters is to choose the proper scale and measures. Only by large time and space scale could the geographical laws be effectively revealed. In this empirical analysis, the study area covers the whole mainland of China, and the time is in 2000. The data include urban population, distance by railway between capital cities, and freight traffic flows of railway between regions. The sources of data is as follows: (1) the urban population is from the 5th population census of China in 2000; (2) the railway distance between cities comes from the mileage table of *Atlas of China Transportation*; and (3) the freight traffic flow of railway between regions is from 2001 *Year Book of China Transportation and Communications*.

Except Taiwan and the two Special Administrative Regions, Hongkong and Macao, there are 31 administrative divisions, including 22 provinces, 5 autonomous regions, and 4 municipalities directly under China's Central Government (Figure 1). In a broad sense, an autonomous region is a province in Chinese. All the provinces, autonomous regions and municipalities are regarded as "regions" in this paper. Among these provinces and municipalities, railroad has not been available to Tibet (autonomous region) and Hainan (province) by 2000. If we adopt the data after 2000, there is no available renewed national census data. In this instance, we decide to exclude Tibet and Hainan from the data set in our analysis. As a result, only 29 regions are taken into account, and the distance and flow data of these regions can form into $29 \times 29$ matrices.

The selection and processing of the data is described as below. First of all, traffic flow data for attraction measure is the most crucial among all data sources, because it determines the values of $F_{ij}$ in equation (1). This kind of data is very difficult to obtain in Chinese mainland, but it's of great significance for the gravity model analysis. Fortunately, the *Year Book of China Transportation and Communication* provides complete railway freight traffic flow data between provinces and municipalities every year.

The second problem is how to measure region sizes by choosing population census data. This set of data is related to the definition of $P_i$ and $P_j$ in equation (1). Generally, urban and regional sizes could be measured by population, output value, and area, etc. Population is the best size measure for China's urban and regional systems. Regional area is constant, and regional output statistics in China have some shortcomings at present. Comparatively speaking, it is more proper to employ population to measure the sizes of cities or regions. China's latest population census is the 5th national census in 2000. As for municipalities, we can use their population census data to quantify their sizes. The problem is how to define the size of a province or an autonomous region. Experimentation shows that it is appropriate to use the total population of systems of cities to respresent the size of a region. The reason is that the attraction measure is in fact limited to freight traffic flow of railway which happened only between cities. If we take the total population (including rural population) of the province or autonomous region or only the population of the provincial capital as the size measure, the result is either too big or too small.

Zhou and Yu (2004) have calculated the population size of each China's city by using the 5th census data, which covers 666 cities. These 666 cities can be divided into 31 groups according to administrative regions. Adding the urban population numbers of cities in each group yields the population size of each region. For example, there are 34 cities in Hebei Province, whose total urban population is 11,498,105 in 2000. So, the size of Hebei Province is defined as $P_{HB}$=11,498,105. Therefore, the size measure of a region can be given by the following formula

$$P_i = \sum_k u_{ik} \; , \tag{16}$$

where $u_{ik}$ represents the census data of the $k$th city in the $i$th province or autonomous region

($i$=1,2,…31; $k$=1,2,…). This definition may not be perfect, but, based on the railway freight flow, no better definition can be found than it for the time being.

The third problem is how to measure the distances between two regions. Theoretically, we could use the distance between two centroids of regions to represent $r_{ij}$ in equation (1). The problem is that this kind of definition is far from reality. An easy but proper method is to adopt the kilometrage by railroad between two capitals of administrative divisions as distance (Figure 1). This method has two advantages. (1) There is ready-made data in *Atlas of China Transportation* that can be used directly; (2) This distance is consistent in track with the freight traffic flow of railway, which is of great importance for our study.

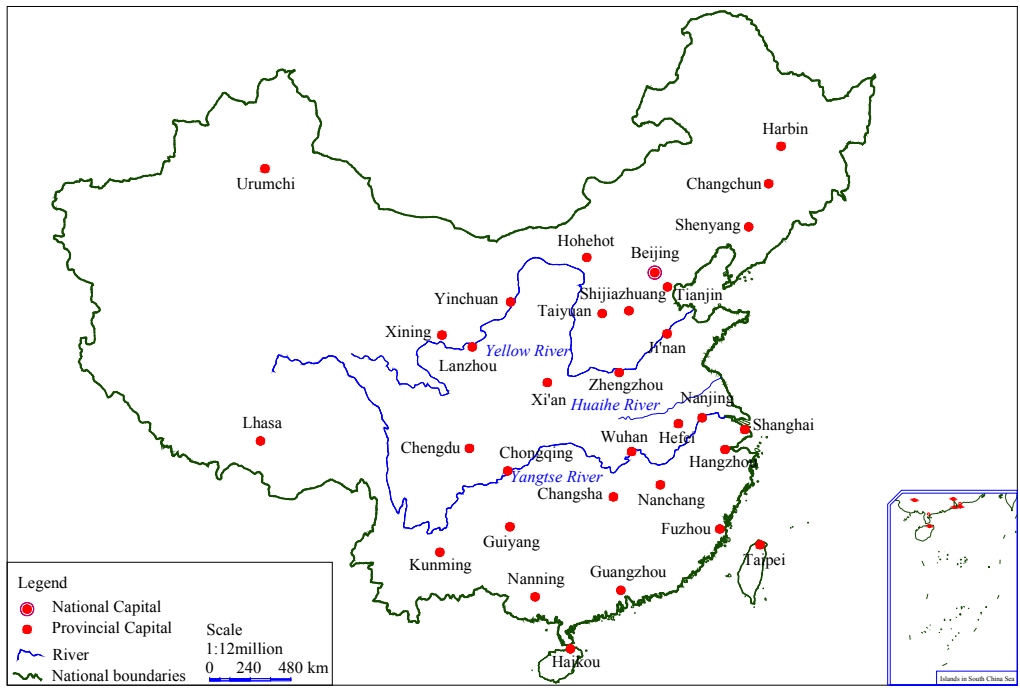

**Figure 1** A simple map of Chinese provincial capitals and municipalities

Through the above data preparation, we can estimate the parameters of the gravity models, equations (2) and (3), by the ordinary least squares (OLS) method, which is practical and simple. After taking logarithms of equations (2) and (3), parameters $α$, $β$, $γ$ and $λ$ change to the slope of linear equations, or what is called regression coefficients. The reason why we make use of OLS method is that it has advantages in estimating the slope of a linear regression model. Other methods, say, the maximum likelihood method may get a more accurate $G$ value, but it can not give solid values of parameters $α$, $β$, $γ$ and $λ$. In geographical systems, the slope is more important than the intercept either in theoretical analysis or in practical use.

## 3.2 Calculation results

For the two forms of urban gravity models as shown in the previous section, which one does the interregional traffic flow of China accord with? People used to judge a mathematical model by certain statistical standards. However, from modern view, such a simple treatment is deficient in many cases. A new point is that a mathematical model can be regarded as good or bad rather than right or wrong (Gabaix and Ioannides, 2004). We should not judge the rights and wrongs of a model only according to whether or not it passes certain statistical tests. Statistical standard only reflects how much the model fits the reality. According to this idea, the following parts will focus on comparing the goodness of fit of the gravity model based on the inverse power function with that based on the negative exponential function. We don't assert which is right and which is wrong.

The method for model-building of spatial interaction demands no more than simple algebra. Preparing for the computation, we should make a linear transform of the gravity models. Taking the logarithm on both sides of equation (2) yields

$$\ln F_{ij} = \ln G + \alpha \ln P_i + \beta \ln P_j - \gamma \ln r_{ij}. \tag{17}$$

Similarly, taking the logarithm of equation (3) gives

$$\ln F_{ij} = \ln G + \alpha \ln P_i + \beta \ln P_j - \lambda r_{ij}. \tag{18}$$

In equations (17) and (18), $P_i$, $P_j$ represent the total population of regions $i$ and $j$ respectively, $r_{ij}$ denotes the distance by railway between the two capitals of the regions, $F_{ij}$ is determined by the freight traffic flow of railway. Then, we can use the method of multivariate linear regression analysis to estimate the values of parameters $\ln G$, $\alpha$, $\beta$, $\gamma$ and $\lambda$. In the following, three kinds of attraction measures will be employed to evaluate these parameters.

**(I) Attraction measure based on traffic inflow or outflow**

In the first set of modeling, we use traffic inflow or outflow as attraction measure by defining

$$F_{ij} = O_{ij}^2 \quad \text{or} \quad F_{ij} = D_{ij}^2. \tag{19}$$

To begin with, we consider the squared value of traffic outflow as attraction measure by taking $F_{ij}=O_{ij}^2$. Based on equation (17), a least square calculation gives all the results of regression coefficients, which are displayed in tables 1 and 2. After a simple transformation, we can get the

following model

$$F_{ij} = O_{ij}^2 = 0.00001571 \frac{P_i^{0.92078} P_j^{1.35809}}{r_{ij}^{2.39925}}, \tag{20}$$

The average value of parameters $\alpha$ and $\beta$ is 1.13943, which is close to 1. The scaling exponent indicating distance friction can be defined by

$$b = \frac{\gamma}{(\alpha+\beta)/2} = \frac{2.39925}{(0.92078+1.35809)/2} = 2.10565. \tag{21}$$

Extracting the 1.13943th root of equation (20) on both sides yields a "standardized" model

$$I_{ij} = F_{ij}^{2/(\alpha+\beta)} = 0.00006082 \frac{P_i^{0.80810} P_j^{1.19190}}{r_{ij}^{2.10565}}. \tag{22}$$

It can be seen that the power exponents of $P_i$ and $P_j$ are close to 1 after transformation, and the sum of the powers is 2.

By using equation (18), a least square calculation gives the following gravity model

$$F_{ij} = O_{ij}^2 = (2.44132 \times 10^{-12}) P_i^{0.93683} P_j^{1.37414} e^{-0.00135 r_{ij}}, \tag{23}$$

The chief regression analysis results are also listed in tables 1 and 2. After extracting the root by the average value of $\alpha$ and $\beta$, namely 1.15549, on both sides of equation (23), the model could be "standardized" in the form

$$I_{ij} = F_{ij}^{2/(\alpha+\beta)} = (8.91708 \times 10^{-11}) P_i^{0.81077} P_j^{1.18923} e^{-0.00117 r_{ij}}. \tag{24}$$

After the standardization process, the sum of the power indexes of $P_i$ and $P_j$ is equal to 2.

**Table 1** Summary output and partial ANOVA results by the attraction measure of traffic outflow square

| Model types | Goodness of fit ($R^2$) | Std error | Total df | F Stat | Significance |
|---|---|---|---|---|---|
| Power | 0.52269 | 1.99316 | 811 | 294.94107 | 2.823E-129 |
| Exponential | 0.46870 | 2.10287 | 811 | 237.59775 | 1.698E-110 |

**Note**: ANOVA is the abbreviation of "**an**alysis **of va**riance", Std error is the ab. of "**St**andar**d error**", df is the ab. of "**d**egree of **f**reedom", and stat is the shortened form of "statistic".

**Table 2** Regression coefficients and related statistics by the attraction measure of traffic outflow square

| Model types | Parameters | | | Stat summary | | Multi-collinearity stat | |
|---|---|---|---|---|---|---|---|
| | Items | Coefficients | Std Error | t Stat | P-value | Tolerance | VIF |
| Power | Intercept | -11.06118 | 2.43320 | -4.54593 | 6.306E-06 | | |
|   | ln$r$ | -2.39925 | 0.11540 | -20.79163 | 3.066E-77 | 0.96874 | 1.03227 |
|   | ln$P_1$ | 0.92078 | 0.09251 | 9.95371 | 4.273E-22 | 0.98231 | 1.01801 |
|   | ln$P_2$ | 1.35809 | 0.09251 | 14.68102 | 1.984E-43 | 0.98231 | 1.01801 |
| Exponential | Intercept | -26.73848 | 2.28594 | -11.69694 | 2.619E-29 |   |   |
|   | $r$ | -0.00135 | 0.00008 | -17.49999 | 2.190E-58 | 0.96546 | 1.03577 |
|   | ln$P_1$ | 0.93683 | 0.09769 | 9.59030 | 1.057E-20 | 0.98056 | 1.01983 |
|   | ln$P_2$ | 1.37414 | 0.09769 | 14.06698 | 2.315E-40 | 0.98056 | 1.01983 |

In the next place, we consider the traffic inflow square as attraction measure, namely take $F_{ij}=D_{ij}^2$. The inflow-based models have no essential difference with the outflow-based models, i.e., equations (20) and (23). The only difference is that the parameters $\alpha$ and $\beta$ exchange their positions in the equations. The models based on inflow can be respectively expressed in the following forms

$$F_{ij} = D_{ij}^2 = 0.00001571 \frac{P_i^{1.35809} P_j^{0.92078}}{r_{ij}^{2.39925}}, \quad (25)$$

$$O_{ij} = D_{ij} = (2.44132 \times 10^{-12}) P_i^{1.37414} P_j^{0.93683} e^{-0.00135 r_{ij}}. \quad (26)$$

**(II) Attraction measure based on the sum of traffic inflow or outflow**

In the second set of modeling, we take the square of the sum of inflow and outflow as attraction measure by defining

$$F_{ij} = (O_{ij} + D_{ji})^2. \quad (27)$$

First, we examine the inverse power law based gravity model. We can get all the regression coefficients of equation (17) by the OLS method, and the results are shown in tables 3 and 4. After simple transformation, the model is expressed as follows

$$F_{ij} = (O_{ij} + D_{ji})^2 = 0.0005627 \frac{P_i^{1.11525} P_j^{1.11525}}{r_{ij}^{2.55215}}. \quad (28)$$

A least square computation yields the parameter values $\alpha=\beta=1.11525$, which are close to 1. Extracting the $\alpha$th root on both sides of equation (28) gives the "standardized" model

$$I_{ij} = (O_{ij} + D_{ji})^{1/\alpha} = 0.0012193 \frac{P_i P_j}{r_{ij}^{2.28840}}. \quad (29)$$

This time, the scaling exponent can be defined by

$$b = \frac{\gamma}{\alpha} = \frac{\gamma}{\beta} = \frac{2.55215}{1.11525} = 2.28840. \tag{30}$$

Second, we try the negative exponential function based gravity model. The OLS computation of equation (18) yields the following results

$$F_{ij} = (O_{ij} + D_{ji})^2 = (3.51125 \times 10^{-11}) P_i^{1.13031} P_j^{1.13031} e^{-0.00145 r_{ij}}, \tag{31}$$

The main regression analysis results are also displayed in tables 3 and 4. After extracting the $\alpha$th root on both sides ($\alpha=\beta=1.13031$), equation (31) could be "standardized" as

$$I_{ij} = (O_{ij} + D_{ji})^{1/\alpha} = (5.63266 \times 10^{-10}) P_i P_j e^{-0.00128 r_{ij}}. \tag{32}$$

As indicated above, the sum of the power indexes of $P_i$ and $P_j$ is equal to 2 after standardization.

**Table 3** Summary output and partial ANOVA results by the attraction measure of the squared sum of traffic inflow and outflow

| Model types | Goodness of fit ($R^2$) | Std error | Total df | $F$ Stat | Significance |
|---|---|---|---|---|---|
| Power | 0.60463 | 1.72407 | 811 | 411.88331 | 2.736E-162 |
| Exponential | 0.54116 | 1.85730 | 811 | 317.65769 | 3.414E-136 |

**Table 4** Regression coefficients and related statistics by the attraction measure of the squared sum of traffic inflow and outflow

| Model types | Parameters | | Stat summary | | | Multi-collinearity stat | |
|---|---|---|---|---|---|---|---|
| | Items | Coefficients | Std Error | $t$ Stat | $P$-value | Tolerance | VIF |
| Power | Intercept | -7.48280 | 2.10470 | -3.55528 | 0.00039952 | | |
| | ln$r$ | -2.55215 | 0.09982 | -25.56859 | 4.048E-106 | 0.96874 | 1.03227 |
| | ln$P_1$ | 1.11525 | 0.08002 | 13.93766 | 1.0014E-39 | 0.98231 | 1.01801 |
| | ln$P_2$ | 1.11525 | 0.08002 | 13.93766 | 1.0014E-39 | 0.98231 | 1.01801 |
| Exponential | Intercept | -24.07247 | 2.01898 | -11.92307 | 2.642E-30 | | |
| | $r$ | -0.00145 | 0.00007 | -21.25003 | 5.959E-80 | 0.96546 | 1.03577 |
| | ln$P_2$ | 1.13031 | 0.08628 | 13.10084 | 1.068E-35 | 0.98056 | 1.01983 |
| | ln$P_1$ | 1.13031 | 0.08628 | 13.10084 | 1.068E-35 | 0.98056 | 1.01983 |

**(III) Attraction measure based on the product of traffic inflow or outflow**

In the third set of modeling, we take the product of inflow and outflow as attraction measure by defining

$$F_{ij} = O_{ij} \times D_{ji}. \tag{33}$$

First of all, we investigate the inverse power law based gravity model. The OLS calculation yields the regression coefficients of equation (17) and related statistics which are listed in tables 5 and 6. Simple transformation gives the following model

$$F_{ij} = O_{ij}D_{ji} = 0.00001571 \frac{P_i^{1.13944} P_j^{1.13944}}{r_{ij}^{2.39925}}, \qquad (34)$$

The parameters are $\alpha=\beta=1.13944$ which are close to 1. Extracting the $\alpha$th root on both sides of equation (34) gives the "standardized" model such as

$$I_{ij} = (O_{ij}D_{ji})^{1/\alpha} = 0.00006082 \frac{P_i P_j}{r_{ij}^{2.10565}}. \qquad (35)$$

In this instance, the coefficient of distance friction can be defined by

$$b = \frac{\gamma}{\alpha} = \frac{\gamma}{\beta} = \frac{2.39925}{1.13944} = 2.10565, \qquad (36)$$

which is close to 2. By comparing equation (35) with equation (22), we can find that the main parameter values of the two models are the same as one another.

Then, we test the negative exponential function based gravity model. A regression analysis of equation (18) gives the following result

$$F_{ij} = O_{ij}D_{ji} = (2.44132 \times 10^{-12}) P_i^{1.15549} P_j^{1.15549} e^{-0.00135 r_{ij}}, \qquad (37)$$

The principal results are also listed in tables 5 and 6. After extracting the root by the value of $\alpha=\beta=1.15549$ on both sides, equation (37) can be changed to the standard form

$$I_{ij} = (O_{ij}D_{ji})^{1/\alpha} = (8.91708 \times 10^{-11}) P_i P_j e^{-0.00117 r_{ij}}. \qquad (38)$$

Comparing equation (38) with equation (24) shows that the main parameter values of the two models are also generally identical to each other.

**Table 5** Summary output and partial ANOVA results by the attraction measure of the product of traffic inflow and outflow

| Model types | Goodness of fit ($R^2$) | Std error | Total df | F Stat | Significance |
|---|---|---|---|---|---|
| Power | 0.61672 | 1.63330 | 811 | 433.37575 | 9.802E-168 |
| Exponential | 0.55216 | 1.76552 | 811 | 332.06673 | 1.917E-140 |

**Table 6** Regression coefficients and related statistics by the attraction measure of the product of traffic inflow and outflow

| Model | Parameters | Stat summary | Multi-collinearity stat |
|---|---|---|---|

| types | Items | Coefficients | Std Error | t Stat | P-value | Tolerance | VIF |
|---|---|---|---|---|---|---|---|
| Power | Intercept | -11.06118 | 1.99390 | -5.54751 | 3.928E-08 | | |
| | $\ln r$ | -2.39925 | 0.09456 | -25.37254 | 6.486E-105 | 0.96874 | 1.03227 |
| | $\ln P_1$ | 1.13944 | 0.07580 | 15.03119 | 3.274E-45 | 0.98231 | 1.01801 |
| | $\ln P_2$ | 1.13944 | 0.07580 | 15.03119 | 3.274E-45 | 0.98231 | 1.01801 |
| Exponential | Intercept | -26.73848 | 1.91922 | -13.93196 | 1.068E-39 | | |
| | $\ln r$ | -0.00135 | 0.00006 | -20.84384 | 1.509E-77 | 0.96546 | 1.03577 |
| | $\ln P_2$ | 1.15549 | 0.08201 | 14.08883 | 1.806E-40 | 0.98056 | 1.01983 |
| | $\ln P_1$ | 1.15549 | 0.08201 | 14.08883 | 1.806E-40 | 0.98056 | 1.01983 |

From the three sets of calculations and results, conclusions can be reached as follows. First, there is no essential difference between the inflow-based results and those based on traffic outflow measure. The only difference lies in that the power exponents of the two size measures exchange their positions. Second, in the average sense, defining the product of inflow and outflow as an attraction measure has the same effect as only defining inflow or outflow as attraction measure. Though the inflow is not equal to the outflow where each region is concerned, the total inflow equals total outflow of all the regions. Third, if we take the sum of inflow and outflow as attraction measure, there is no intrinsic difference between the calculation results of the exponential-type gravity model based on different attraction measures, except some estimation errors of the model's parameters. However, there are obvious differences between the computation results of the power-type gravity model based on different attraction measures. As a whole, taking the product of traffic inflow and outflow as attraction measure, as stipulated by equation (6), is more proper and easier for us to build models and make analysis. The last but not least, the scaling exponent indicative of distance friction is about $b$=2.1, which is very close to the value of the power exponent of the law of gravity ($b$=2) in classical physics. This result is in agreement with the theoretical expectation value derived from the fractal theory of urban hierarchy and network.

## 4 Questions and discussions

### 4.1 Model forms and parameters

The data of Chinese freight traffic flow of railway can be fitted to the two kinds of the gravity models, equation (2) and equation (3), or equation (7) and equation (8). Comparatively speaking, the effect of fit of the power-law-based model is better than that of the exponential-based model.

Although both the inverse power function and the negative exponential function can reflect the distance-decay effect in geographical space, the physical meanings is much different (table 7). The inverse power function reflects action at a distance while the negative exponential function suggest spatial locality (Chen, 2008). In the negative exponential function, there exists a parameter indicating characteristic length, i.e., $\lambda=1/r_0$. If the distance $r$ is greater than the mean distance $r_0$, the force of attraction will attenuate to zero rapidly. In contrast, the inverse power function is scale-free, or of scaling invariance. There is no characteristic length for the distance variable in the inverse power-law distribution.

Autocorrelation analysis and spectral analysis can show the difference between these two kinds of impedance functions. The autocorrelation function (ACF) of the inverse power function attenuates much slower than that of the negative exponential function. The partial autocorrelation function (PACF) of the inverse power function decays more slowly than that of the negative exponential function does. The PACF of the negative exponential function displays a sharp cutoff at a displacement of 1 (Chen, 2008). On the other hand, the wave-spectrum relation of the inverse power function implies complex spatial interaction, while the spectral exponent of the negative exponential function suggests only weak spatial interaction, say, a geography unit can only affect its immediate units but has no direct effect on the alternate units (Chen and Zhou, 2008; Liu and Chen, 2007). As for the Chinese system of regions, the spatial interaction indicates not so much locality as action at a distance, as is shown by the statistics in tables 1 to 6.

The form of the impedance function of the gravity model gives rise to many controversies over different opinions. A large mount of research has been made on the gravity models, yet a great many studies are no guarantee of considerable progress. The inverse power-law based gravity model was once criticized and even abandoned for its dimensional problem while the negative exponential based gravity model brings new problems, especially locality dilemma. According to the spectral analysis, if we take the negative exponential function as impedance function, the spectral exponent is close to 2 indicating local spatial interaction (Chen and Zhou, 2008), which contradicts the action at a distance that gravity model demands. In fact, the negative exponential function implies simple structure while the inverse power function suggests complex structure (Barabasi and Bonabeau, 2003; Chen and Zhou, 2004). Since cities and regions are both complex spatial systems (Allen, 1997; Wilson, 2000), the impedance function of the gravity model is more likely to be the inverse power law. In the past, based on idea from Euclidean geometry, geographers could not explain the dimension of the inverse power-law based gravity model (Haynes, 1975). Nowadays, owing to the application of fractal theory to geography, the dimensional problem associated with the distance friction coefficient is no longer a theoretical puzzle. In fact, both of the negative exponential based gravity model and the inverse power law based gravity model have their own scopes of application. The former is applicable inside a metropolitan area or at smaller spatial scale, while the latter is applicable in a region or at larger spatial scale. The former is related with Euclidean geometrical patterns, while the latter is related to fractal geometrical patterns (see table 7).

**Table 7** Comparison between the gravity model based on the inverse power law and that based on the negative exponential function

| Items | Power-law based gravity model | Exponential-based gravity model |
| --- | --- | --- |
| Spatial interaction | Action at a distance | Locality |
| Physical feature | Complex patterns | Simple patterns |
| Geometrical structure | Fractal structure | Euclidean geometrical structure |
| Application range | Interurban area, large scale | Intraurban area, small scale |

In all kinds of attraction measures, it seems to be the best to employ the function of the product of inflow ($D_{ji}$) and outflow ($O_{ij}$), $(O_{ij}D_{ji})^{1/\alpha}$, as a measure of attraction force, $F_{ij}$. The parameter $\alpha$ is determined by the corresponding regression coefficient. As for China's case in 2000, $\alpha$ is about 1.139. We conjecture that $\alpha$ might be 1 in the ideal geographical condition. But this is only a conjecture at present. It needs further theoretical demonstration and empirical evidences to substantiate this judgment. In terms of the analysis of traffic flows between China's regions, it is realistic and satisfying to take equation (6) as the definition of attraction measure. Whether for the inverse power law based gravity model or for the negative exponential based one, the mode to quantify the attraction force are the same.

It should be noted that the calculation process in this paper is different from other geographers' method. As we know, the matrix of spatial distance between regions is symmetric. However, the table of traffic flows is an asymmetric matrix as the inflow and outflow between two regions are always unequal to one another. Geographers used to employ only the upper or lower triangle matrix to estimate the gravity parameters. This kind of approach is not accurate. The reasonable method is to take the whole matrix of spatial distances and that of traffic flows to evaluate the gravity parameters. For the gravity model based on the inverse power function, we should remove the diagonal elements of the distance matrix and the traffic flow matrix; but for the gravity model based on the negative exponential function, the entire elements of the matrices can be adopted.

The gravity models of interregional flows in China are made at the macro level and the effect is desirable. For the power-law-based model, the scaling exponent is around $b$=2.1, which consists with that given by Yang (1990) before (see also Wang, 2001). Another conjecture we have is that the distance friction coefficient under the ideal geography conditions is very close to $b$=2. The reasons why the results given by some geographers used to be far from 2 may rest with two aspects. One is owing to the data processing method, for instance, only the upper or lower triangle matrixes were taken into consideration; the other is because that the sample sizes were not large enough so that the results were unstable.

The parameters of the gravity model have puzzled the geographers for a long time. Decades ago, theoretical geographers tried in vain to derive the parameters of a gravity model from theoretical probability distributions (Harris, 1964; Harvey, 1971; Lowry, 1964; Schneider, 1959). Despite no expected results, these works are instructive for us to understand the nature and properties of human geographical systems deeply. Maybe there are no constants for the urban gravity model, and the parameters change over space and time (Mikkonen and Luoma, 1999). The variability of the parameters reflects the asymmetry of geographical laws, which further reveal the complexity and nonconservation of geographical systems. It is of great significance for us to study the nonconservation of geographical systems and the asymmetry of geographical laws in order to apprehend and interpret the nature of the human geographical phenomena.

**4.2 Distance-decay function**

There is a closely inherent relation between the gravity models and the distance attenuation function. The distance-decay analyses can supplement the gravity analyses. We can examine the distance-decay relationships between Chinese regions based on the traffic inflow or outflow. From equations (6), (7) and (8), we can derive the following distance-decay functions

$$O_i = O_1 r_i^{-b}, \tag{39}$$

$$O_i = O_0 e^{-b r_i}, \tag{40}$$

where $r_i$ and $O_i$ denote the distance and traffic outflow from the original place to destination place $i$ respectively, and $b$, $O_0$, and $O_1$ are all parameters. Taking the distances as independent variable and the traffic outflow as dependent variable, we can make a regression analysis. The result shows that the outflows of most regions follow the inverse power function, equation (39), a few cases satisfy the negative exponential function, equation (40), and the rest fail to show any clear distance-decay patterns. For example, the traffic outflow from Beijng ($O$), the capital of China, can be fitted to equation (39), and the scatter points and trend line are displayed in figure 2(a).

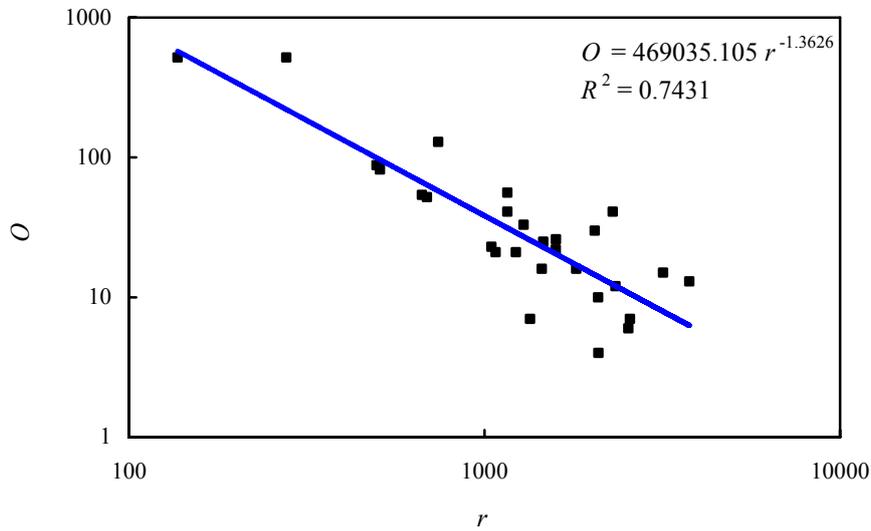

a. Beijing as an origin of outflow

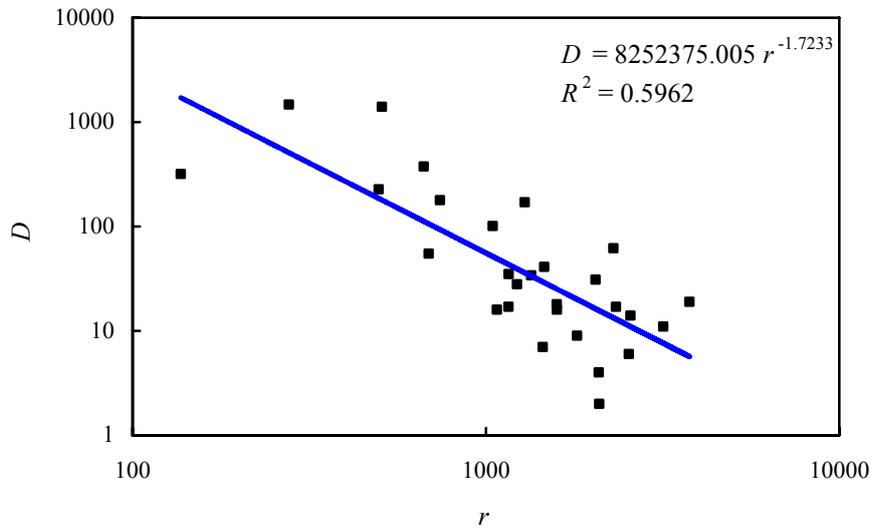

b. Beijing as a destination of inflow

**Figure 2** The power-law distributions of the traffic inflows and outflows of Beijing (2000)

Further, taking the distance from the destination place to original place *j* as independent variable and the inflow from original place *j* to the destination region as independent variable, we can also compare the fitting results of the following two models

$$D_j = D_1 r_j^{-b}, \tag{41}$$

$$D_j = D_0 e^{-br_j}. \tag{42}$$

where $D_j$ refers to the traffic inflow from original place *j* to the destination place, $D_0$ and $D_1$ are parameters, the remaining notation is the same as in equations (39) and (40). The result shows that the inflows of most regions follow the inverse power law, equation (41), a few meet the negative exponential distribution, equation (42), and the rest are not very clear. Taking the municipality of Beijng as a destination (*D*), its distance attenuation pattern of traffic inflow is shown in figure 2(b).

The great majority of the results of China's cities and regions lend more support to the power law based gravity model than that to the exponential function based gravity model.

### 4.3 Gravity model and spatial optimization index

Because of the asymmetry of human geography laws, there are essential differences between the geographical models and the classical physical models. For classical physics, a counterexample is enough to falsify a possible law. But for human geography, a counterexample is not sufficient to negate a possible law. On the other hand, for classical physics, reality is the supreme evidence that tests a model. But for human geography, not only can reality test a model but also the model can test reality in turn. When a model does not fit the geographical system well, it is possible that the model is right but reality is 'wrong'—this seemingly extreme but in fact profound view comes from August Lösch. In short, we can evaluate the distance between actual states and optimal state of a geographical system by the models which proved to be appropriate for many good cases. As for the gravity model, we can construct a formula to quantify the distance from the optimization, and the expression is

$$h = \frac{1}{2}(\frac{\alpha}{\alpha^*} + \frac{b}{b^*}) - 1, \tag{43}$$

where *h* can be termed as *optimization deviation index* (ODI) reflecting the extent far from optimization, $\alpha$ denotes the power exponent of size measures of the gravity model, $\alpha^*$ represents the exponent of size measures under the ideal condition, $b$ refers to the distance friction coefficient, $b^*$ represents the distance friction coefficient under the ideal condition. Equation (43) suggest that the closer the parameters of the gravity model are to the standard values, the closer the value of $h$ is to 0, and the shorter the distance from actual state to the optimal state is. We don't consider the coefficient $G$ in the model just because that the intercept in equations (17) and (18) is far less important than the slopes. In the great majority of cases, the gravity coefficient values in different models for different geographical systems lack comparability.

Based on the third sets of calculation results, it is not difficult to figure out the ODI values of spatial distribution of the freight traffic flows of railway between China's administrative regions. For the inverse power based gravity model, we have $\alpha^*=1$, $b^*=2$, thus

$$h = \frac{1}{2}(1.13944 + \frac{2.10565}{2}) - 1 = 0.096 . \tag{44}$$

For the negative exponential based gravity model, we have $\alpha^*=1$, $b^*=1/r_0$, where $r_0$ represents the average distance between provincial capitals. The average distance betweens the 29 regions is about 1780.589 km, whose reciprocal $b^*$ is around 0.00056, so we have

$$h = \frac{1}{2}(1.15549 + \frac{0.00117}{0.00056}) - 1 = 0.622 . \tag{45}$$

The ODI of China's interregional traffic flows given by the inverse power based gravity model is 0.096, while the ODI determined by the negative exponential based gravity model is 0.622. The latter is much greater than the former. This result implies that: (1) the inverse power law based gravity model is more suitable for the distribution of China's interregional flows, and (2) the small ODI value given by the inverse power based gravity model suggests that China's interregional flows distributions is not far from the optimized state. The results of spatial optimization analyses give further support to the power-law based gravity model.

## 5 Conclusions

The attraction measure, scaling exponent, and impedance function of the urban gravity model are redefined in this article. All the theoretical results and the mathematical derivations are supported by the empirical analysis of China's systems of cities and regions. The main original work made in this paper can be summarized as follows.

First, the attraction measure of spatial interaction in human geographical systems is defined as a function of the product of traffic inflow and outflow. As an approximate substitute, the attraction measure can also be defined as the squared inflow or squared outflow. The key of estimating the parameter values of the gravity model is to make reasonable use of the whole elements of the flow matrix rather than only the upper or lower triangle matrix. Based on the new definition of attraction measure, the urban gravity model stemming from physical analogy with the Newton's law of universal gravitation is demonstrated to be different from Wilson's spatial interaction model deriving from entropy-maximization principle.

Second, the scaling exponent of the power law based gravity model is demonstrated to be the ratio of the fractal dimension of network of cities to that of hierarchies of cities. The fractal dimension of hierarchies of cities ($D$) is theoretically equal to that of city-size distribution which approaches to unity ($D=1$); the fractal dimension of network of cities ($D_f$) approaches to two if the geographical space is fully filled up by cities ($D_f \to 2$). As a result, the value of the scaling exponent indicating distance friction approaches to two ($b=D_f/D \to 2$), which corresponds to the value of the power exponent of the law of gravity in classical physics.

Third, the inverse power function as an impedance function of the gravity model is more acceptable than the negative exponential function on the whole. A theoretical hypothesis is that the gravity models based on different impedance functions have different scopes or spheres of application. The inverse power law indicates complexity, structure, and action at a distance, therefore the gravity model based on the power law is suitable for large scale of regions or complex urban systems. In contrast, the negative exponential law implies simplicity, randomicity, and locality, thus the gravity model based on the exponential function is suitable for small scale of regions and relatively simple urban systems. Anyway, we agree with the following viewpoint that, for the congener but different models, we can say which is better and which is not better for a concrete object in question, but we cannot say which is right and which is wrong.

**Acknowledgement**: This research was sponsored by the National Natural Science Foundation of China (Grant No. 40771061) and Beijing Natural Science Foundation (Grant No. 8093033).